\documentclass[11pt]{article}
\usepackage[final]{coling}
\usepackage{times,amsmath,amssymb,latexsym,graphicx,multirow,bm,url,footmisc}
\usepackage[T1]{fontenc}
\usepackage[utf8]{inputenc}
\usepackage{microtype}
\usepackage{inconsolata}

\title{DiffStyleTTS: Diffusion-based Hierarchical Prosody Modeling for 
Text-to-Speech with Diverse and Controllable Styles}

\author{
 \textbf{Jiaxuan Liu\textsuperscript{1}},
 \textbf{Zhaoci Liu\textsuperscript{1}},
 \textbf{Yajun Hu\textsuperscript{2}},
 \textbf{Yingying Gao\textsuperscript{3}},
 \textbf{Shilei Zhang\textsuperscript{3}},
 \textbf{Zhenhua Ling\thanks{Corresponding author.
 This work was funded by the National Nature Science Foundation of China under Grant U23B2053.}\textsuperscript{1}},
\\
 \textsuperscript{1}NERCSLIP, University of Science and Technology of China, Hefei, China,
\\ 
\textsuperscript{2}iFLYTEK CO.LTD., Hefei, China,
\\
 \textsuperscript{3}China Mobile Research Institute, Beijing, China
\\
 \texttt{\{jxliu, zcliu8\}@mail.ustc.edu.cn, yjhu@iflytek.com,}\\
 \texttt{\{gaoyingying, zhangshilei\}@chinamobile.com, zhling@ustc.edu.cn}
}
\begin{document}
\maketitle
\begin{abstract}
  Human speech exhibits rich and flexible prosodic variations. 
  To address the one-to-many mapping problem from text to prosody in a reasonable and flexible manner, 
  we propose DiffStyleTTS, a multi-speaker acoustic model based on a conditional diffusion module and an
  improved classifier-free guidance, 
  which hierarchically models speech prosodic features, 
  and controls different prosodic styles to guide prosody prediction.
  Experiments show that our method outperforms all baselines in naturalness and
  achieves superior synthesis speed compared to three diffusion-based baselines. 
  Additionally, by adjusting the guiding scale, 
  DiffStyleTTS effectively controls the guidance intensity of the synthetic prosody.
\end{abstract}

\section{Introduction}

Speech synthesis, also known as text-to-speech (TTS), 
aims to turn text into almost human-like audio. 
Currently, most TTS models consist of three main components: 
a text analysis front-end, an acoustic model, and a vocoder.
Among them, the naturalness and prosodic performance of speech primarily 
depend on the design of the acoustic model.

The acoustic model, at the heart of TTS, 
can be categorized as autoregressive and non-autoregressive. 
Autoregressive acoustic models, like Tacotron \cite{Tacotron} and Transformer TTS \cite{TransformerTTS}, 
have issues with word skipping, repeated reading and inference time increasing linearly with 
length of the Mel-spectrogram. 
Non-autoregressive acoustic models, like FastSpeech2 \cite{FastSpeech2}, excel in rapidly synthesizing high-quality speech. 
However, they are constrained by using a simple regression objective function for optimization, 
lacking probabilistic modeling, and the unimodal characteristics of Gaussian distribution 
don't conform to the true distribution of acoustic features, which affects the prediction accuracy. 
The mean of the distribution also results in the problem of over-smoothing predictions, 
which restricts the diversity of generated prosodic features. 
These issues lead to weak fluctuations and unnaturalness in prosodic transfer and control tasks, 
Additionally, traditional prosodic transfer methods like Global Style Tokens (GST) \cite{GST} lack controllability over the intensity of prosodic transfer.

The recently emerged diffusion model has significant advantages 
in describing the complex distribution of high-dimensional and multi-modal features. 
In particular, the guidance of the conditional diffusion model \cite{CG} can well control the results. 
It effectively addresses issues like over-smoothing predictions and a lack of diversity through multi-step sampling. 
Currently, acoustic models based on the diffusion model, 
such as Diff-TTS \cite{Diff-TTS}, Grad-TTS \cite{Grad-TTS}, DiffSinger \cite{DiffSinger}, Guided-TTS \cite{Guided-TTS}, ProDiff \cite{ProDiff}, 
CoMoSpeech \cite{CoMoSpeech}, etc., 
primarily use the Mel-spectrogram as the prediction target. 
There have been limited studies on predicting speech prosodic features via the conditional diffusion model.
DiffProsody\cite{DiffProsody} is a diffusion-based prosody prediction model that 
constrains prosodic features through a discriminator, 
but it still lacks controllability over prosody during inference.
In summary, the flexible transfer and control of speech prosody still remains underexplored.

Therefore, we propose a novel acoustic model, DiffStyleTTS, 
based on a conditional diffusion module and an improved classifier-free guidance \cite{CFG}. 
It hierarchically models prosodic features using both coarse-grained style conditions and fine-grained prosodic descriptions, 
balances the diversity and quality of prosody via classifier-free guidance, 
and is applied in prosodic transfer and control tasks to guide prosody prediction. 
Additionally, we introduce the dynamic thresholding method to address the issue of phoneme distortion caused by an excessive guiding scale.
DiffStyleTTS is also designed to support various inference modes.
Experiments \footnote{\url{https://xuan3986.github.io/DiffStyleTTS/}\label{demo}} show DiffStyleTTS achieves higher naturalness and similar or faster synthesis speed
compared to FastSpeech2 and other diffusion-based baselines. 
Compared to the FastSpeech2+GST and DiffProsody baselines, it demonstrates superior prosodic transfer capability, 
enabling flexible combination of speaker and prosodic features, 
alongside controllable prosodic transfer intensity via a guiding scale.

\section{DiffStyleTTS}
\label{sec:DiffStyleTTS}
\begin{figure*}[tb]
	\centering
	\centerline{\includegraphics[width=\linewidth]{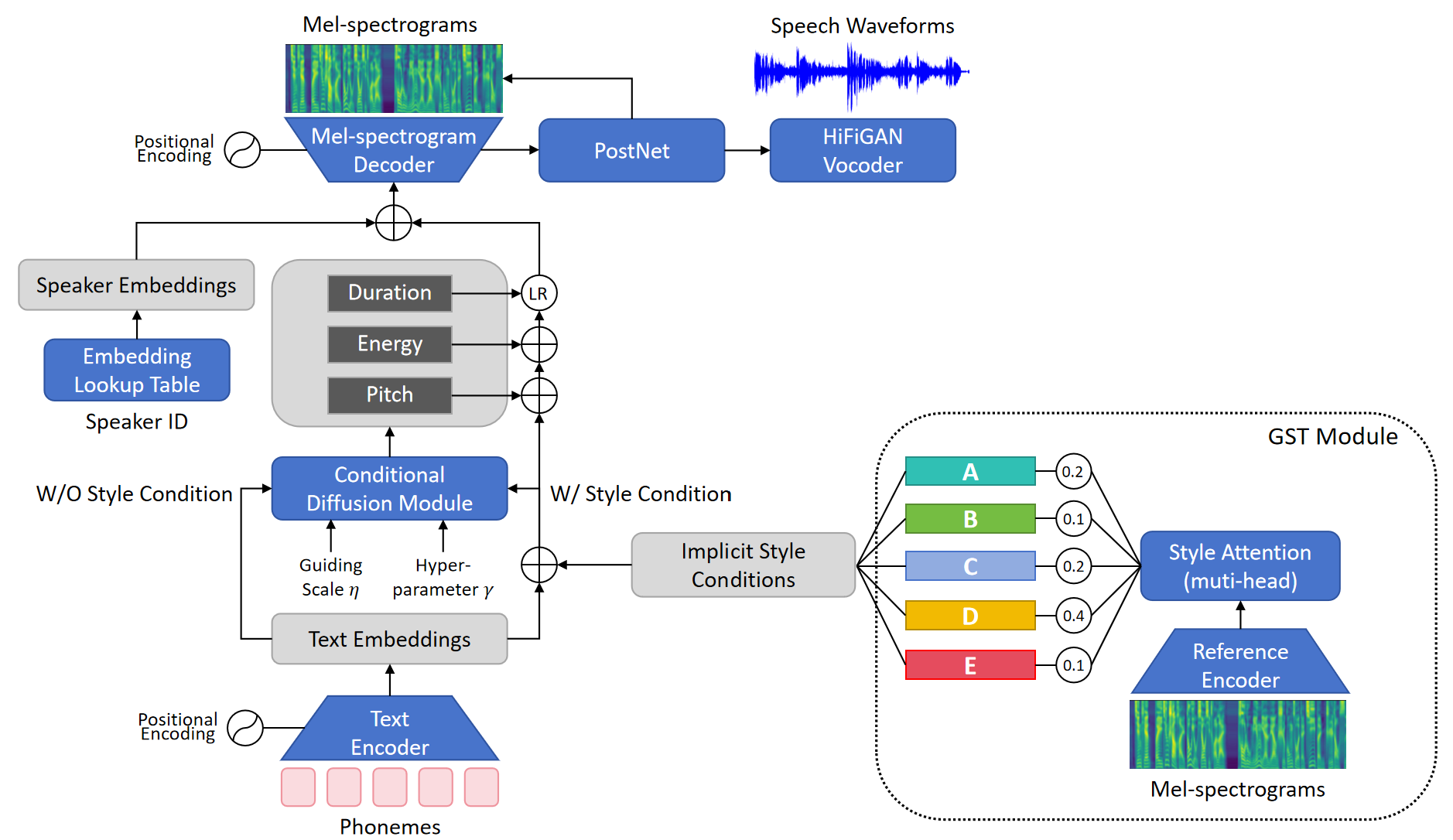}}
	\caption{The model architecture of DiffStyleTTS. The LR refers to length regulator. The conditional diffusion module includes two denoisers $\varPsi_{\theta_1}(\bm{x}_t,t,\bm{y},\bm{c})$ and $\varPsi_{\theta_2}(\bm{x}_t,t,\bm{y})$.}
	\label{fig:DiffStyleTTS}
\end{figure*}

In this section, we propose DiffStyleTTS, a multi-speaker acoustic model that employs hierarchical prosody modeling 
and utilizes FastSpeech2 as its backbone.
As shown in Figure \ref{fig:DiffStyleTTS}, the encoder and decoder use the feed-forward Transformer
(FFT) of FastSpeech2, along with a 5-layer convolutional PostNet \cite{Tacotron2} in the decoder. 
We use an embedding lookup table to capture the unique vocal characteristics of each speaker 
and a HiFi-GAN vocoder \cite{HiFi-GAN} to synthesize speech waveforms. 
The main modifications are replacing FastSpeech2's original variance adaptor with a conditional diffusion module for hierarchical prosody modeling and introducing a GST module for style control.

\subsection{Hierarchical Prosody Modeling}
\label{ssec:Hierarchical Prosody Modeling}
The DiffStyleTTS achieves hierarchical prosody modeling by considering prosodic features at two levels: 
coarse-grained implicit style conditions and fine-grained explicit prosodic descriptions. 
Implicit style conditions encompass broad descriptions of entire sentences, 
which are difficult to define intuitively and are encoded from the Mel-spectrogram 
during the training of the whole acoustic model.
Explicit prosodic features include fine-grained prosodic descriptions of phonemes, 
such as pitch, energy, and duration, which can be directly and easily extracted from speech waveforms. 

In DiffStyleTTS, the method of GST \cite{GST} is adopted to 
extract implicit style conditions from audio as shown in Figure \ref{fig:DiffStyleTTS}. 
The implicit style conditions are decoupled into style vectors corresponding to a group of global style tokens.
Furthermore, explicit prosodic features are predicted using a conditional diffusion module, 
which introduces text embeddings and implicit style conditions as its conditional terms. 

\subsection{The Conditional Diffusion Module}
\label{ssec:The Conditional Diffusion Module}
The conditional diffusion module is guided with implicit style conditions to predict explicit prosodic features that align with it.
The guided generation of the conditional diffusion module can be usually categorized into two approaches: 
classifier guidance \cite{CG} and classifier-free guidance \cite{CFG}.
Classifier guidance requires an additional classifier, 
which slows down the inference speed, and its quality impacts the effectiveness of category generation. 
Therefore, DiffStyleTTS employs classifier-free guidance, which can avoid these issues as it doesn't require direct calculation of the classifier gradient.

\begin{figure}[tb]
	\centering
	\centerline{\includegraphics[width=\columnwidth]{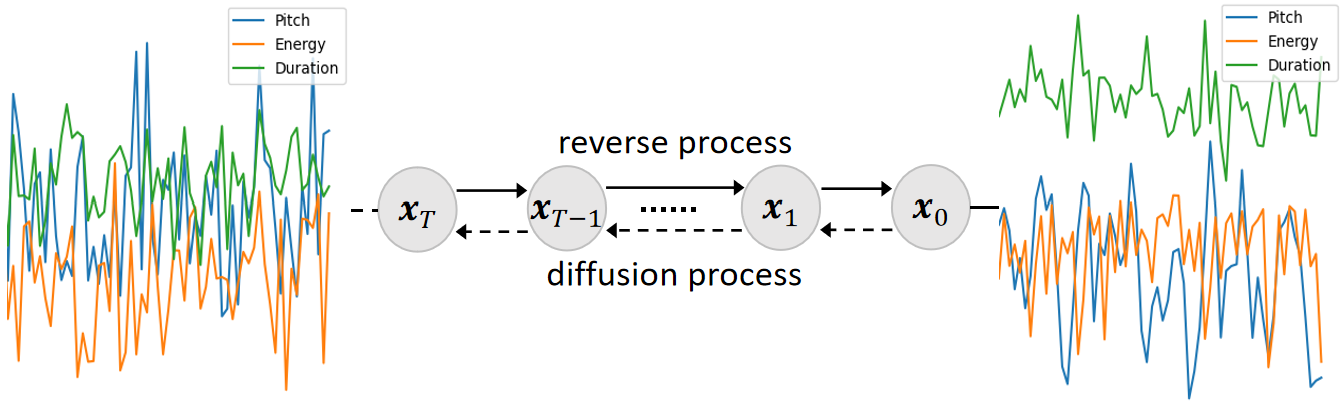}}
	\caption{The diffusion process and reverse process of explicit prosodic features.}
	\label{fig:diffusion}
\end{figure}

First, as shown in Figure \ref{fig:diffusion}, based on the mathematical principles of DDPM \cite{DDPM},
the diffusion process of explicit prosodic features is defined by a fixed Markov chain from the initial data \bm{$x_0$} to the latent variable \bm{$x_t$} as
\begin{gather}
q(\bm{x}_{1:T}|\bm{x}_0)=\prod_{t=1}^Tq\left(\bm{x}_t|\bm{x}_{t-1}\right),\\
q(\bm{x}_t|\bm{x}_{t-1})=\mathcal{N}(\bm{x}_t;\sqrt{1-\beta_t}\bm{x}_{t-1},\beta_t\mathbf{I}),
\end{gather}
where $\bm{x}_t=\sqrt{\bar{\alpha}_t}\bm{x}_0+\sqrt{1-\bar{\alpha}_t}\bm{\epsilon}$, $\alpha_t=1-\beta_t$, $\bar{\alpha}_t=\prod_{s=1}^t\alpha_s$, 
$t=0,1,\cdots,T$, and $T$ is the step size. 
When adding a small Gaussian noise at each step, 
the module selects a small positive constant $\beta_t$ from a variance table, 
which we define as a cosine schedule \cite{IDDPM} to prevent rapid noise accumulation from linear addition.
Then, the reverse process is also defined by a Markov chain from $\bm{x}_t$ to $\bm{x}_0$ parameterized by the $\theta $ as
\begin{gather}
p_\theta(\bm{x}_{0:T})=p(\bm{x}_T)\prod_{t=1}^Tp_\theta(\bm{x}_{t-1}|\bm{x}_t),\\
\hspace{-1.5mm}
\!p_\theta(\bm{x}_{t-1}|\bm{x}_t)\!=\!\mathcal{N}\big(\bm{x}_{t-1};\bm{\mu}_\theta(\bm{x}_t,t),\bm{\Sigma}_\theta(\bm{x}_t,t)\big),\!
\end{gather}
which shows the step-by-step denoising of an isotropic Gaussian noise $\bm{x}_T\thicksim\mathcal{N}(\bm{0},\mathbf{I})$ to restore the original data $\bm{x}_0$.

To guide the conditional diffusion module's output using classifier-free guidance, 
two denoisers with identical architectures are designed to employ text embeddings $\bm{y}$ as a condition 
to learn the mapping of phonemes into explicit prosodic features, with and without implicit style conditions $\bm{c}$ respectively. 
The denoiser $\varPsi_{\theta_1}(\bm{x}_t,t,\bm{y},\bm{c})$ uses $\bm{y}$ and $\bm{c}$ 
as the conditional input ($\bm{condition}=\bm{y}+\bm{c}$), 
while the denoiser $\varPsi_{\theta_2}(\bm{x}_t,t,\bm{y})$ leaves $\bm{c}$ empty, 
using only $\bm{y}$ as the conditional input.
At each denoising step, the input of each denoiser consists of explicit prosodic features $\bm{x}_t$ linearly combined with 
noise $\bm{\epsilon}\thicksim\mathcal{N}(\bm{0},\mathbf{I})$. 
To model this noise, two denoisers are trained using the following training objective
\begin{gather}
	\label{formula:5}
	\begin{aligned}
	min_{\theta_1}L_{diff\_c}(\theta_1)&=\\
	\mathbb{E}_{\bm{\epsilon},\bm{x}_t,t,\bm{y},\bm{c}}&\parallel\bm{\epsilon}-\bm{\epsilon}_{\theta_1}(\bm{x}_t,t,\bm{y},\bm{c})\|_2^2,
	\end{aligned}\\
	\label{formula:6}
	\begin{aligned}
	min_{\theta_2}L_{diff\_nc}(\theta_2)&=\\
	\mathbb{E}_{\bm{\epsilon},\bm{x}_t,t,\bm{y}}&\parallel\bm{\epsilon}-\bm{\epsilon}_{\theta_2}(\bm{x}_t,t,\bm{y})\|_2^2,
	\end{aligned}
\end{gather}
where we get noise outputs $\bm{\epsilon}_{\theta_1}(\bm{x}_t,t,\bm{y},\bm{c})$ and $\bm{\epsilon}_{\theta_2}(\bm{x}_t,t,\bm{y})$.

During inference, the two noise outputs are linearly interpolated to obtain the guided results
\begin{equation}
	\begin{aligned}\tilde{\bm{\epsilon}}_{\theta_1,\theta_2}(&\bm{x}_t,t,\bm{y},\bm{c})=\bm{\epsilon}_{\theta_2}(\bm{x}_t,t,\bm{y},\varnothing)+\\
	&\eta\big(\bm{\epsilon}_{\theta_1}(\bm{x}_t,t,\bm{y},\bm{c})-\bm{\epsilon}_{\theta_2}(\bm{x}_t,t,\bm{y},\varnothing)\big).\end{aligned}
\end{equation}
Here, $\eta$ is the guiding scale used to adjust the guidance intensity, 
balancing the diversity and quality of explicit prosodic features.

Preliminary experiments found that when $\eta$ was too high ($\eta\geq7.0$), phoneme distortion occasionally occured in some phonemes. 
These phonemes exhibited noise or elongation phenomena, 
resembling the ``overexposed'' issue from the previous study \cite{overexposed}. To fix this, 
we improve classifier-free guidance via dynamic thresholding method, correcting the standard deviation of guidance results at each sampling step
\begin{gather}
	\begin{aligned}
		\sigma_{cond}=&std\big(\bm{\epsilon}_{\theta_1}(\bm{x}_t,t,\bm{y},\bm{c})\big),\\
		\sigma_{cfg}=&std\big(\tilde{\bm{\epsilon}}_{\theta_1,\theta_2}(\bm{x}_t,t,\bm{y},\bm{c})\big),
	\end{aligned}\\
	\hspace{-1.5mm}
	\!\tilde{\bm{\epsilon}}_{rescaled}(\bm{x}_t,t,\bm{y},\bm{c})\!=\!\tilde{\bm{\epsilon}}_{\theta_1,\theta_2}(\bm{x}_t,t,\bm{y},\bm{c})\!\cdot\!\frac{\!\sigma_{cond}\!}{\!\sigma_{cfg}\!},\!\\
	\begin{aligned}
		\hspace{-2mm}
		\tilde{\bm{\epsilon}}_{final}=\gamma\tilde{\bm{\epsilon}}_{rescaled}&(\bm{x}_t,t,\bm{y},\bm{c})+\\
		&(1-\gamma)\tilde{\bm{\epsilon}}_{\theta_1,\theta_2}(\bm{x}_t,t,\bm{y},\bm{c}).
	\end{aligned}
\end{gather}
This method corrects the standard deviation of $\tilde{\bm{\epsilon}}_{\theta_1,\theta_2}(\bm{x}_t,t,\bm{y},\bm{c})$ to 
the original standard deviation of $\bm{\epsilon}_{\theta_1}(\bm{x}_t,t,\bm{y},\bm{c})$.
A correction scale $\gamma$ adjusts the intensity of the correction to achieve the final corrected result $\tilde{\bm{\epsilon}}_{final}$.

\begin{figure}[tb]
	\begin{minipage}[b]{\linewidth}
		\centering
		\centerline{\includegraphics[width=\columnwidth]{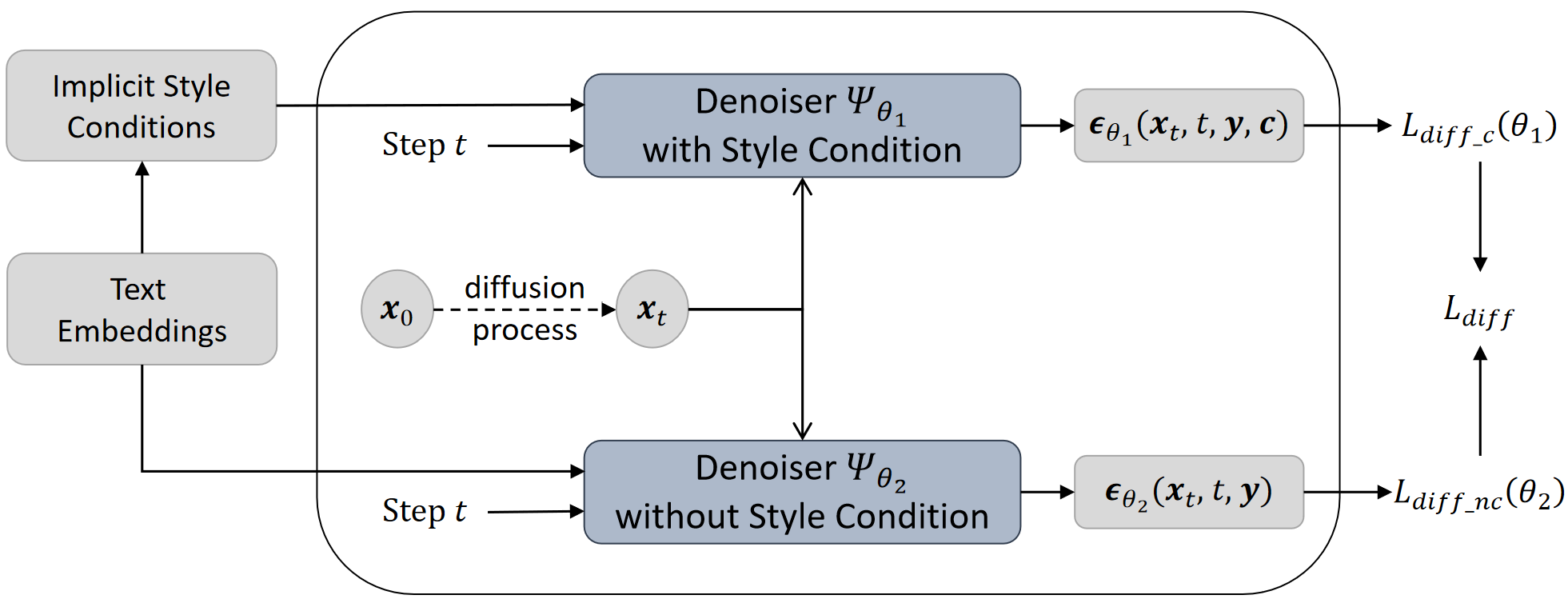}}
		\centerline{(a) Training process.}\medskip
	\end{minipage}
	\hfill
	\begin{minipage}[b]{\linewidth}
		\centering
		\centerline{\includegraphics[width=\columnwidth]{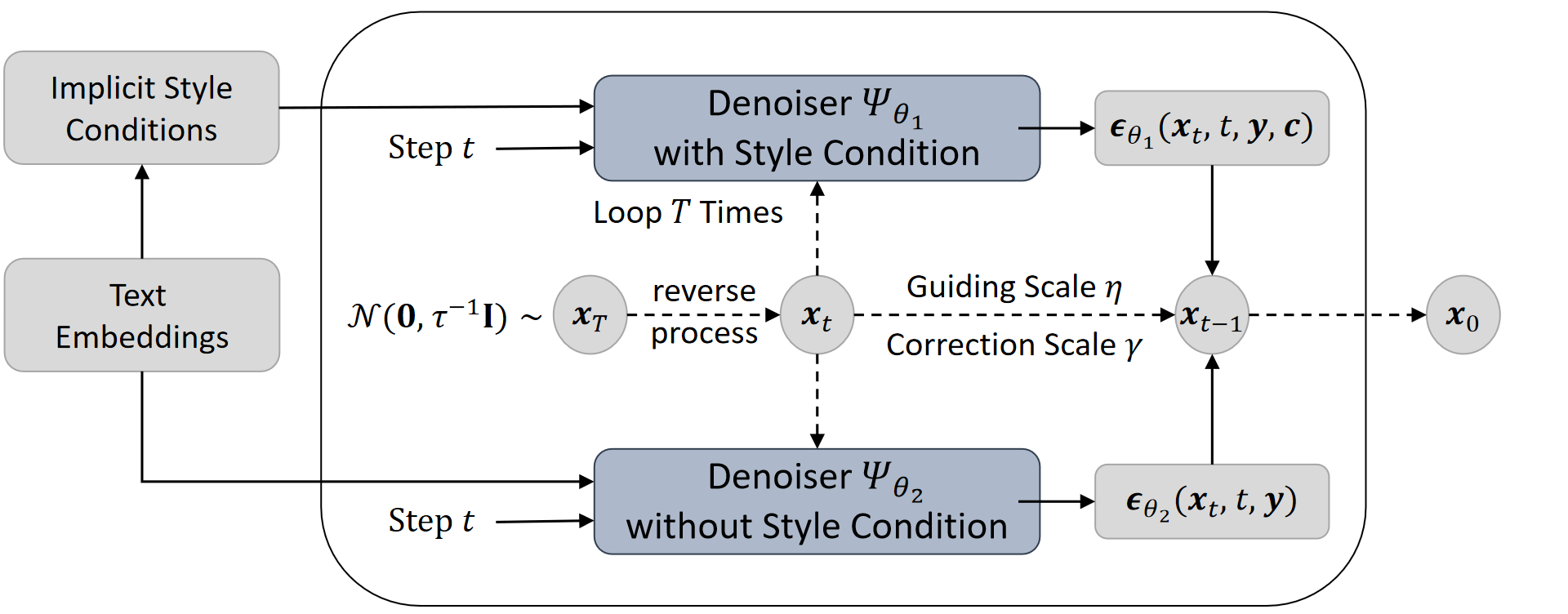}}
		\centerline{(b) Inference process.}
	\end{minipage}
	\caption{An illustration of the training and inference processes of the conditional diffusion module based on classifier-free guidance.}
	\label{fig:CFG}
\end{figure}

In summary, the conditional diffusion module can employ implicit style conditions $\bm{c}$ 
to guide the generation of explicit prosodic features,
and use the guiding scale $\eta$ and the correction scale $\gamma$ to flexibly adjust the diversity of the explicit prosodic features 
and the guidance intensity.

\subsection{Training and Inference}
\label{ssec:Training and Inference}

\subsubsection{Training}
\label{sssec:Training}
Referring to Figure \ref{fig:DiffStyleTTS} and Figure \ref{fig:CFG}(a), processed phonemes from the text analysis front-end 
are fed into the text encoder to generate text embeddings. 
This, along with implicit style conditions from the GST module, is then input into the conditional diffusion module,
which includes two denoisers $\varPsi_{\theta_1}(\bm{x}_t,t,\bm{y},\bm{c})$ and $\varPsi_{\theta_2}(\bm{x}_t,\bm{t},\bm{y})$ trained via Eq. (\ref{formula:5})(\ref{formula:6}).
During training, the log scales of raw phoneme-wise pitch and duration, and raw phoneme-wise energy, are targeted for sampling. 
The guiding scale $\eta$ and the correction scale $\gamma$ are not involved in the training process.
We add the implicit style conditions and the embeddings of raw pitch and energy to the text embeddings, 
then employ the length regulator to align the length based on raw duration. 
Frame-wise speaker embeddings are added before feeding into the decoder. 
Finally, decoded Mel-spectrograms are converted into speech waveforms using the pre-trained HiFi-GAN vocoder. 
The total loss function includes the diffusion module loss, the loss of decoding Mel-spectrograms, and 
the residual loss of PostNet
\begin{equation}
	\begin{aligned}
	L_{total}=L_{diff\_c}(\theta_1)+&L_{diff\_nc}(\theta_2)+\\
	&L_{decoder}+L_{mel}.
	\end{aligned}
\end{equation}

\subsubsection{Inference}
\label{sssec:Inference}
Referring to Figure \ref{fig:DiffStyleTTS} and Figure \ref{fig:CFG}(b), 
three main inference modes are designed based on the trained DiffStyleTTS model.

(1) \textit{Diversified controllable inference.} By tuning the guiding scale $\eta$ and the correction scale $\gamma$, 
we can adjust the diversity and guidance intensity of explicit prosodic features,
achieving diversified and controllable prosody prediction.

(2) \textit{Prosodic transfer inference.} Given a reference utterance and a specified speaker ID, 
prosodic features are transferred from the reference utterance to this speaker. 
By tuning the guiding scale $\eta$ and the correction scale $\gamma$, 
we can adjust the intensity of prosodic transfer.

(3) \textit{Prosodic control inference.} Given a specified speaker ID and a token ID in the GST module, 
we can set the weights of other tokens to 0 and the weight of this token to 1, 
synthesizing prosody controlled only by that token. 
Besides, we allow for the flexible combination of style token weights, 
enabling the enhancement or diminishment of certain prosodic style. 
We can also scale the pitch, energy, and duration by multiplication with scaling factors 
to control prosodic values.

Additionally, we introduce a temperature hyperparameter $\tau$ \cite{Grad-TTS} to sample terminal condition $\bm{x}_T$ from $\mathcal{N}(\bm{0},\tau^{-1}\mathbf{I})$ 
instead of $\bm{x}_T$ from $\mathcal{N}(\bm{0},\mathbf{I})$. Previous work has found that tuning $\tau$ can help to improve the quality of output. 

\section{Experiments}
\label{sec:Experiments}

\subsection{Experimental Setup}
\label{ssec:Experimental Setup}

\subsubsection{Dataset}
\label{sssec:Dataset}
We evaluated the proposed DiffStyleTTS model using a 54-hour private Mandarin Chinese dataset 
comprised of recordings from 9 male speakers of different ages, all of which belonged to the genres of novel, 
narration or story reading. Our dataset has high recording quality and diverse prosodic styles. 
We randomly sampled 20 utterances from each speaker's recordings, and the total 180 utterances were reserved for validation and test, 
while the rest were used for training.
All phoneme durations were extracted by an internal forced alignment tool based on HMM.
Given phoneme boundaries, phoneme-wise pitch and energy features were obtained by averaging frame-wise pitch and energy. 
The frame-wise pitch was extracted using STRAIGHT \cite{STRAIGHT} and interpolated at unvoiced frames, 
and the frame-wise energy was computed as the L2-norm of the amplitude spectrum derived using short-time Fourier transform. 
To get the input of the text encoder, each phoneme was represented as the concatenation of a phoneme identity embedding, a tone
embedding and a positional embedding.

\subsubsection{Model Configuration}
\label{sssec:Model Configuration}
The encoder encoded phonemes to 256-D text embeddings using 4 FFT blocks.
while the decoder used 6 FFT blocks. 
The 5-layer convolutional PostNet in the decoder is comprised of 512 filters with shape 5$\times$1
with batch normalization, followed by tanh activations on all but the final layer.
The architecture of two denoisers utilized a bidirectional dilated convolution \cite{DiffWave}
similar to WaveNet \cite{WaveNet}, for predicting waveform signals. 
It consists of a stack of 12 residual layers, each layer with residual channels $C=3$ and
a kernel size of 3.
The input tensor had a shape of $[B, C, L]$, where $B$ was the batch size, $C$ was the residual channels, 
and $L$ was the length of phonemes. 
In the GST module, the token embeddings size was set to 256 and the token size was configured to 10.
the multi-head attention with 4 attention heads used a softmax activation to output weights over the tokens.

\subsection{Performance of Synthetic Speech}
\label{ssec:Performance of Synthetic Speech}
Subjective and objective evaluations were conducted to evaluate the performance of speech synthesized using DiffStyleTTS.
In addition to the FastSpeech2 \cite{FastSpeech2} baseline, 
a FastSpeech2 model with ground truth phoneme-wise prosodic features, a Grad-TTS \cite{Grad-TTS} model 
, a Guided-TTS \cite{Guided-TTS} model and a DiffProsody \cite{DiffProsody} model were also built for comparison. 
First, the naturalness mean opinion scores (MOS) of all models were evaluated by a listening test.
A total of 14 participants evaluated 18 utterances for each model, selecting two samples from each speaker.
Second, the accuracy of predicted prosodic probability distributions were evaluated using Jensen-Shannon (JS) Divergence.
Third, the efficiency of different models were evaluated using the real time factor (RTF).
Fourth, different settings of $\eta$ and $\gamma$ used only for prosodic transfer and control
can affect the MOS of DiffStyleTTS, as shown in Table \ref{table:Diversity}. 
Therefore, we select the optimal configuration in this section.
We employed the \textit{diversified controllable inference} mode with $\eta=1.0$ and $\gamma=0.7$ in DiffStyleTTS.
In all experiments, the step size $T$ of all diffusion models was set to 200 to ensure rigor.
Additionally, we ran multiple trials from different terminal conditions $\bm{x}_T$ to sample various explicit prosodic features, 
and then averaged scores for final evaluation results.

\begin{table}[tb]
	\centering
	\caption{The naturalness MOS, JS Divergence and RTF of different models. 
	The FastSpeech2* refers to FastSpeech2 with ground truth phoneme-wise prosodic features, 
	the ISC refers to implicit style conditions, 
	the TEC refers to text embedding conditions,
	and the AISC refers to adding implicit style conditions into text embeddings.}
	\label{table:MOS}
	\medskip
	\resizebox{\linewidth}{!}{
		\begin{tabular}{l c c c c c}
		\hline
			\multirow{2}{*}{Model}&\multirow{2}{*}{MOS $\uparrow$}&\multicolumn{3}{c}{JS Divergence $\downarrow$}&\multirow{2}{*}{RTF$\downarrow$}\\
		\cline{3-5}
			&&Pitch&Energy&Duration&\\
		\hline
			Ground Truth&4.55$\pm$0.05&-&-&-&-\\
		\hline
			FastSpeech2&3.85$\pm$0.06&0.121&0.037&0.097&0.019\\
			FastSpeech2*&4.11$\pm$0.07&-&-&-&-\\
			Grad-TTS&4.08$\pm$0.07&0.115&0.040&0.088&0.250\\
			Guided-TTS&4.15$\pm$0.07&0.080&0.033&0.050&0.479\\
			DiffProsody&4.10$\pm$0.06&0.083&0.030&0.046&0.063\\
			{\bf DiffStyleTTS}&{\bf 4.18$\pm$0.06}&{\bf 0.065}&{\bf 0.030}&{\bf 0.045}&{0.048}\\
		\hline
			w/o ISC&3.92$\pm$0.07&0.090&0.038&0.078&-\\
			w/o AISC&3.80$\pm$0.06&0.071&0.033&0.051&-\\
			w/o TEC&2.05$\pm$0.24&0.445&0.125&0.390&-\\
		\hline
		\end{tabular}
	}
\end{table}
\begin{table}[tb]
	\centering
	\caption{The naturalness MOS, CV (\%) of different guiding scale $\eta$ in DiffStyleTTS.}
	\label{table:Diversity}
	\medskip
		\begin{tabular}{l c c c c}
		\hline
			\multirow{2}{*}{$\eta$}&\multirow{2}{*}{MOS $\uparrow$}&\multicolumn{3}{c}{CV $\uparrow$}\\
		\cline{3-5}
			&&Pitch&Energy&Duration\\
		\hline
			1.0&4.23$\pm$0.06&7.75&3.04&5.46\\
			3.0&4.09$\pm$0.08&8.52&3.20&4.87\\
			5.0&3.87$\pm$0.07&12.04&4.25&8.69\\
			7.0&3.79$\pm$0.08&16.63&5.81&8.94\\
		\hline
		\end{tabular}
\end{table}
The results are shown in the first six rows of Table \ref{table:MOS}. 
We can see that our proposed DiffStyleTTS outperformed all baselines in naturalness and JS Divergence. 
DiffStyleTTS also achieved faster synthesis speed than the two diffusion models using Mel-spectrograms as modeling targets.

\subsection{Prosodic Control}
\label{ssec:Prosodic Control}
To study the effect of the guiding scale $\eta$ in classifier-free guidance on balancing prosodic diversity and quality, 
we selected an utterance with pronounced prosodic variations 
and synthesized it using the \textit{diversified controllable inference} mode with $\gamma=0.7$.
We used naturalness MOS to evaluate the quality, 
and then calculated the coefficient of variation (CV) of phoneme-wise pitch, energy and duration to evaluate the diversity. 
A total of 12 participants evaluated four values of $\eta$ to illustrate the effect of the guiding scale.
The evaluation results, presented in Table \ref{table:Diversity}, 
indicate that as the $\eta$ increases, the diversity of explicit prosodic features increases 
while the audio quality decreases. 
Notably, when $\eta\geq7.0$, phoneme distortion occasionally occured.

\begin{figure}[tb]
	\centering
	\centerline{\includegraphics[width=\columnwidth]{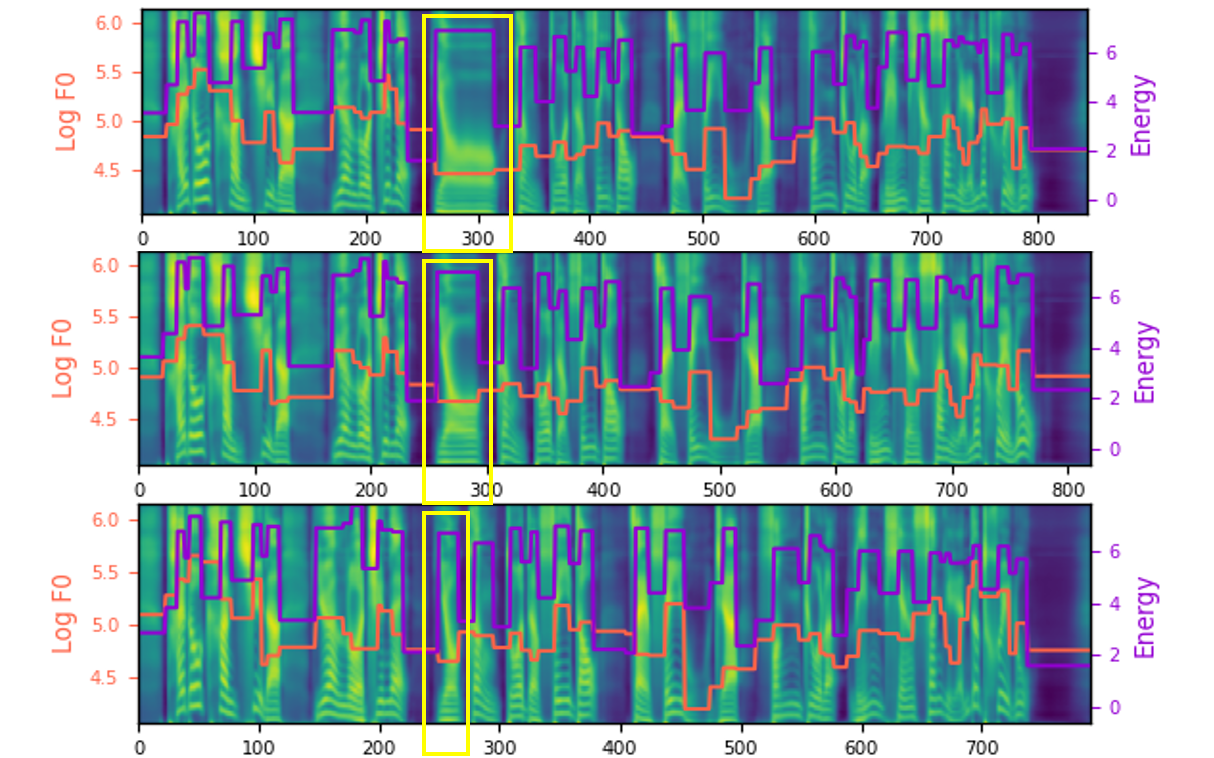}}
	\caption{The Mel-spectrograms of a sentence synthesized with $\gamma=0$ (top), 0.4 (middle) and 0.7 (bottom) respectively.}
	\label{fig:gama}
\end{figure}

\begin{figure}[tb]
	\centering
	\centerline{\includegraphics[width=\columnwidth]{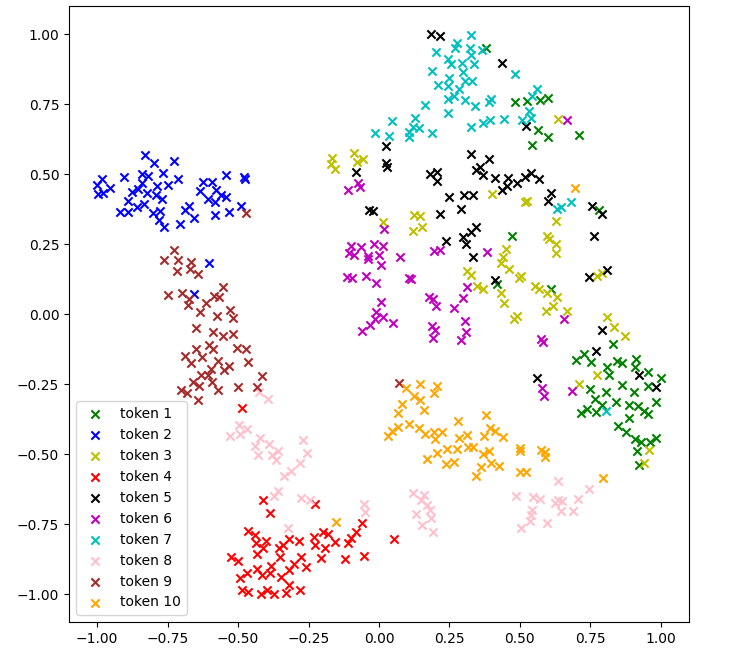}}
	\caption{The t-SNE visualization results.}
	\label{fig:tsne}
\end{figure}

To study the effect of the correction scale $\gamma$ in the dynamic thresholding method, 
we randomly selected a test sentence, set the guiding scale $\eta=7.0$ to cause phoneme distortion,
and employed the \textit{diversified controllable inference} mode to synthesize it three times with different values of $\gamma$.
Figure \ref{fig:gama} shows a distorted phoneme within the yellow box. 
When $\gamma$ was set to 0 (top), the phoneme distortion was pronounced, exhibiting noticeable elongation.
This issue was alleviated at $\gamma=0.4$ (middle) and effectively resolved at $\gamma=0.7$ (bottom).

To verify the ability of different tokens in the GST module to control the generation of explicit prosodic features, 
we randomly selected 50 sentences from the evaluation set to predict explicit prosodic features. 
During inference, we employed the \textit{prosodic control inference} mode  
to synthesize a total of 500 explicit prosodic samples, which included phoneme-wise pitch, energy, and duration. 
We then computed the mean, standard deviation, median, minimum, maximum, skewness, and kurtosis for each sample, 
resulting in a 21-dimensional prosodic distribution vector for each sample. 
We used t-SNE \cite{t-SNE} to reduce the dimensions of these vectors to two dimensions for visualization.

Figure \ref{fig:tsne} shows that these 500 samples can be well divided into 10 clusters, 
each matching one of the 10 tokens. 
which indicates the effectiveness of the hierarchical prosody modeling in DiffStyleTTS 
on controlling the prosodic features of synthetic speech via implicit style conditions.

\subsection{Prosodic Transfer}
\label{ssec:Prosodic Transfer}
\begin{figure*}[tb]
	\begin{center} 
	\begin{minipage}[b]{0.24\linewidth}
		\centering
		\centerline{\includegraphics[height=2.9cm]{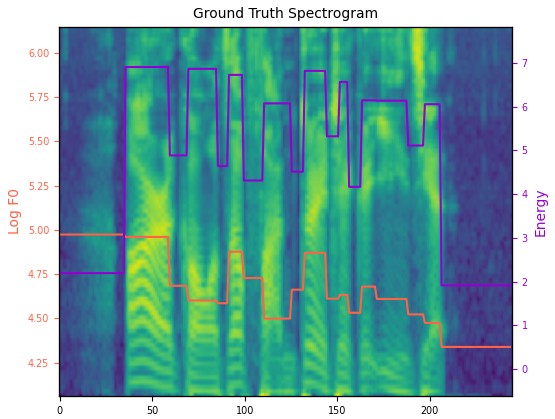}}
		\centerline{(a) Original}
	\end{minipage}\hfill
	\begin{minipage}[b]{0.24\linewidth}
		\centering
		\centerline{\includegraphics[height=2.9cm]{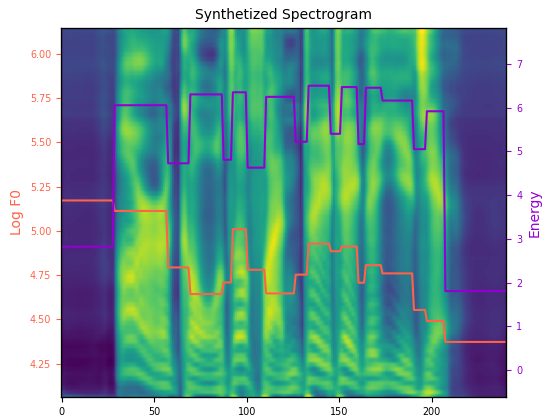}}
		\centerline{(b) $\eta=0.5$}
	\end{minipage}\hfill
	\begin{minipage}[b]{0.24\linewidth}
		\centering
		\centerline{\includegraphics[height=2.9cm]{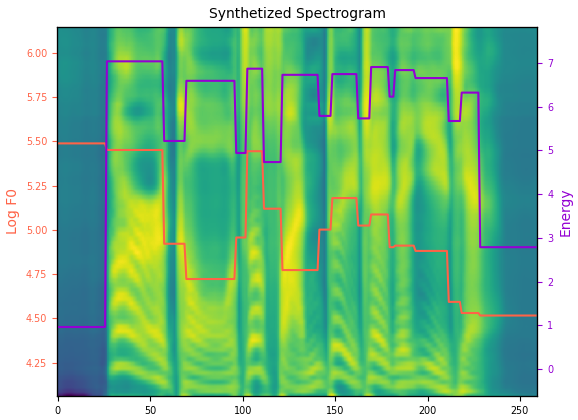}}
		\centerline{(c) $\eta=1.0$}
	\end{minipage}\hfill
	\begin{minipage}[b]{0.24\linewidth}
		\centering
		\centerline{\includegraphics[height=2.9cm]{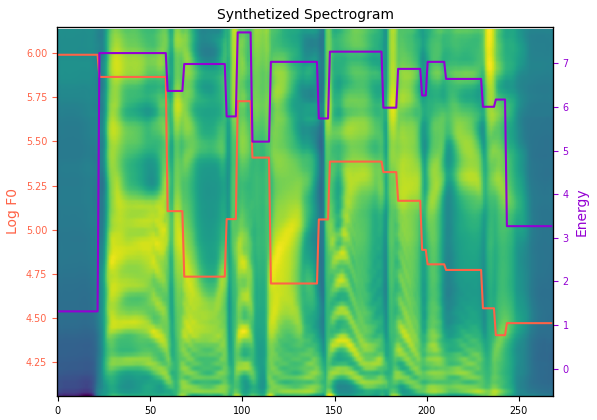}}
		\centerline{(d) $\eta=2.0$}
	\end{minipage}
	\end{center}
	\caption{By tuning the guiding scale $\eta$, 
	we adjusted the intensity of the prosodic transfer from reference audio to original audio (a) to 
	synthesize three Mel-spectrograms (b) (c) and (d).}
	\label{fig:tranfer}
\end{figure*}

To evaluate the effect of DiffStyleTTS in prosodic transfer, 
we used FastSpeech2+GST, i.e., incorporating a GST module before the variance adaptor in FastSpeech2,
and the DiffProsody as the baselines, 
we selected one reference utterance each from two speakers (A and B) with distinct prosodic styles. 
Then, we randomly selected two sentences each from the other eight speakers, excluding the reference speakers, 
resulting in a total of 16 sentences for prosodic transfer. 
The \textit{prosodic transfer inference} mode was used to transfer the prosody of the reference utterance to the other eight speakers.
We conducted a subjective preference (\%) test involving 12 participants to compare the two models and 
calculated the p-value of t-test to assess significance of differences. 
As shown in Table \ref{table:preference}, DiffStyleTTS significantly outperformed two baselines 
on this prosodic transfer task ($p<0.05$).
We also investigated the effect of the guiding scale $\eta$ on the intensity of prosodic transfer, setting $\gamma=0.7$.
The experimental results in Figure \ref{fig:tranfer} compare the transfer effects of three different guiding scales.
It's observed that as $\eta$ increased, the prosodic transfer effect became more pronounced.
We recommend listening to the examples on the demo page \footref{demo}.

\begin{table}[tb]
	\centering
	\caption{Subjective preference results for prosodic transfer.}
	\label{table:preference}
	\resizebox{\linewidth}{!}{
		\begin{tabular}{l c c c c}
		\hline
			\multirow{2}{*}{}&\multicolumn{3}{c}{Preference}&\multirow{2}{*}{p-value}\\
		\cline{2-4}
			&FastSpeech2+GST&Neutral&DiffStyleTTS&\\
		\hline
		    Reference A&33.13&14.37&\bf 52.50&0.0077\\
			Reference B&31.25&21.87&\bf 46.88&0.0249\\
		\hline
		\end{tabular}
	}
	\resizebox{\linewidth}{!}{
		\begin{tabular}{l c c c c}
		\hline
			\multirow{2}{*}{}&\multicolumn{3}{c}{Preference}&\multirow{2}{*}{p-value}\\
		\cline{2-4}
			&DiffProsody&Neutral&DiffStyleTTS&\\
		\hline
		    Reference A&16.67&33.33&\bf 50.00&0.0134\\
			Reference B&18.13&29.37&\bf 52.50&0.0231\\
		\hline
		\end{tabular}
	}
\end{table}

\subsection{Ablation Studies}
\label{ssec:Ablation Studies}
We conducted ablation studies to demonstrate the effectiveness of key components in DiffStyleTTS.
The results are presented in the last three rows of Table \ref{table:MOS}. 

To verify the guiding effect of implicit style conditions on explicit prosodic features, 
we can observe that removing the implicit style conditions $\bm{c}$ from the denoiser $\varPsi_{\theta_1}(\bm{x}_t,t,\bm{y},\bm{c})$,
prohibiting the use of classifier-free guidance, i.e., w/o ISC, but retaining them added to text embeddings, led to a decrease in MOS and an increase in JS divergence. 
When the implicit style conditions added to the text embeddings were removed and only retained as the condition $\bm{c}$ in $\varPsi_{\theta_1}(\bm{x}_t,t,\bm{y},\bm{c})$, 
i.e., w/o AISC, we can also observe a decrease in MOS, which verifies the implicit style conditions contain prosodic features.
Furthermore, removing the text embedding conditions $\bm{y}$ from the denoiser $\varPsi_{\theta_1}(\bm{x}_t,t,\bm{y},\bm{c})$
, i.e., w/o TEC, resulted in poor quality of synthetic prosody, indicating the importance of text embedding conditions on prosodic alignment.
In summary, experiments show that these key components contributed significantly to the performance of DiffStyleTTS.

\section{Conclusion}
\label{sec:illust}
This paper proposes a multi-speaker acoustic model, DiffStyleTTS, 
based on a conditional diffusion module and an improved classifier-free guidance. 
We hierarchically model prosodic features at both implicit and explicit levels. 
Text embeddings and implicit style conditions are combined as the diffusion module's conditions.
To predict explicit prosodic features,
the dynamic thresholding method is employed to improve classifier-free guidance and then adjust the guidance intensity.
Experiments show that our proposed model achieves higher naturalness compared to all baselines,
and faster synthesis speed compared to diffusion-based baselines.
Additionally, DiffStyleTTS demonstrates superior prosodic transfer capabilities and flexibility in comprehensive prosodic transfer and control.

\section{Limitation}
Although our work on prosody prediction has made it more flexible and controllable, we haven't successfully decoupled prosody from speaker timbre.
In addition, we can also observe some overlapping samples after dimensionality reduction, indicating that the tokens are not entirely
independent and share common prosodic styles.
This suggests that the implicit style conditions need further classification.
To achieve better disentanglement of prosody will be a task for our future work.
\bibliography{custom}

\appendix



\end{document}